\documentclass{article}

\usepackage{PRIMEarxiv}

\usepackage[utf8]{inputenc} 
\usepackage[T1]{fontenc}    
\usepackage{url}            
\usepackage{booktabs}       
\usepackage{amsfonts}       
\usepackage{nicefrac}       
\usepackage{microtype}      
\usepackage{lipsum}
\usepackage{fancyhdr}       
\usepackage{graphicx}       
\graphicspath{{media/}}     
\usepackage{amsmath}
\usepackage{xcolor}
\usepackage{multirow}
\usepackage{tabularx,booktabs}

\usepackage{draftwatermark}

\SetWatermarkText{Pre-print}
\SetWatermarkColor[gray]{0.5}
\SetWatermarkFontSize{1cm}
\SetWatermarkAngle{90}
\SetWatermarkHorCenter{20cm}

\title{Impact of Frequency Support by Wind Turbines on
Small-Signal Stability of Power Systems
}

\author{
  Antonio Pepiciello,\\
  University of Sannio, \\
   Piazza Roma 21, 82100, Benevento, Italy.\\
  \texttt{apepiciello@unisannio.it} \\
   \And
  José Luis Domínguez-García \\
  Catalonia Institute for Energy Research, IREC, \\
  Jardins de les Dones de Negre 1, 08930 Sant Adria de Besos, Barcelona, Spain.\\
  \texttt{jldominguez@irec.cat} \\
   \And
     Alfredo Vaccaro,\\
  University of Sannio, \\
   Piazza Roma 21, 82100, Benevento, Italy.\\
  \texttt{vaccaro@unisannio.it} 
}

\begin{document}
\maketitle

\begin{abstract}
Rising wind energy integration, accompanied by a decreasing level of system inertia, requires additional sources of ancillary services. Wind turbines based on doubly fed induction generators (DFIG) can provide inertial and primary frequency support, when equipped with specific controls. This paper investigates the effect of frequency support provision by DFIGs on the small-signal stability of power systems. To this end, a modified version of the Kundur two-area test system is employed to analyze different scenarios. Wind energy generation is either added to the existing system or displaces part of the synchronous generation. Simulations show that primary frequency support tends to improve the damping of electromechanical oscillations and deteriorate it for converter control-based ones. On the other hand, inertial response may be either beneficial, detrimental or negligible to damping, depending on the tuning of control parameters. 
\end{abstract}

\keywords{small-signal stability, wind power generation, doubly fed induction generator (DFIG), inertia, transient stability, oscillation damping.}

Wind energy capacity has been growing exponentially over the last decade, as carbon neutrality became a crucial goal for policy makers worldwide. This radical transition, from centralized and dispatchable power plants to decentralized and stochastic renewable energy sources, poses unprecedented challenges to power systems \cite{ahmed2020grid}. 

Doubly fed induction generators (DFIG) are expected to play a key role in many countries' grid integration targets. However, due to their operating differences with synchronous generators, they raise concerns about power system operation.


The first impact is associated with the location of DFIGs, strictly related to the presence of wind, their primary energy source. Such locations are different from current power plants, resulting in the modification of power flow paths and their resulting synchronizing and damping forces.
More importantly, the reduction of synchronous generation, as wind integration displaces less efficient power plants, involves operating power systems with lower levels of inertia.
Indeed, DFIGs cannot provide natural inertial response, since they are decoupled from the grid by power electronic converters.

Other consequences of large wind integration are the reduced number of Power System Stabilizers (PSS) on the network and less resources providing ancillary services, such as frequency and voltage regulation. 
Furthermore, DFIGs do not sense the fluctuations of frequency caused by a contingency on the network and they do not currently provide the same ancillary services as thermal and hydroelectric power plants \cite{milano2018foundations}.

A possible solution is equipping DFIGs with appropriate controls, for the provision of emulated inertial response and primary frequency support \cite{attya2018review}, as expected by the evolving grid codes worldwide \cite{diaz2014participation}.



The concept of wind turbines providing frequency support was introduced by the authors of \cite{morren2006wind}, which showed how controls based on frequency deviation and Rate of Change of Frequency (ROCOF) can improve the frequency stability of the system. The authors of \cite{chang2009strategies} described how active power from wind turbines can be controlled similarly to synchronous generators. Further works focused either on emulated inertia or primary  frequency response, for example  \cite{ruttledge2015emulated, mahish2019distributed}, providing different control alternatives. Finally, the effect of frequency support provision by wind farms on ROCOF and frequency nadir was assessed in \cite{ruttledge2012frequency, attya2017insights}.


Although it is well-known that the lack of synchronous inertia from wind farms negatively affects the frequency response of the system after a contingency \cite{tielens2016relevance}, the results on small-signal stability analysis and low-frequency oscillation damping are still ambiguous. 

The effect of the dynamic interaction between DFIGs and power systems on small-signal stability is demonstrated by \cite{du2017strong}, which explored the modal coupling of control loop models. According to the same author, the impact of DFIGs on low-frequency oscillation damping depends on the test systems and their operating conditions \cite{du2017small}. For example, the authors of \cite{gautam2009impact} state that the introduction of DFIGs can have either positive or negative effects on damping, depending on the sensitivity to system parameters like power flows, location of the wind farms and load conditions. A discussion on how reactive power control strategies affect oscillation damping is reported in \cite{vittal2011rotor} where the authors conclude that controlling terminal voltage instead of providing a fixed amount of reactive power can be more beneficial in terms of damping. 
In \cite{wilches2016impact}, the effect of wind farm location on inter-area oscillations is explored and it is shown that the changes in damping appear to be negligible, although damping is slightly worsened when the wind farm is in the importing area. In order to distinguish the impact of load flow changes and dynamic interactions introduced by DFIGs, in \cite{du2015method} a method based on the damping torque analysis for the separate examination of these two affecting factors is proposed.
Finally, the analysis carried out by \cite{allen2016measurement}, showed that the effects of DFIGs on small-signal stability can also be assessed by data-driven methods, such as power spectral density analysis.


Equipping DFIGs with PSSs, similarly to synchronous generators, might be a successful strategy to improve oscillation damping  \cite{dominguez2012power}. Different approaches have been proposed: one possibility is taking local signals, from the point of common coupling, as an input to the PSS \cite{surinkaew2014coordinated}. 
Another interesting challenge is the optimal selection of the input signal to PSSs, based on controllability and observability aspects of the signals \cite{dominguez2014input}.

As is clear from this literature review, there is an increasing trend of equipping DFIGs with controls for inertial and primary frequency support provision. Furthermore, additional controls for low-frequency oscillation damping from DFIGs are being proposed. However, little research effort has been devoted to the analysis of the interactions between frequency stability and low-frequency oscillations, which are strictly intertwined. First results on the optimization of DFIG fast frequency response, accounting for small-signal stability constraints, are presented in \cite{huang2021optimization}. Furthermore, \cite{su2012influence} analyzed the impact of frequency support provision by direct-drive-full-converter-based wind farms on small-signal stability and concluded that ancillary frequency control has a beneficial effect on low-frequency oscillation damping.

To the best of the authors’ knowledge, no previous work analyzed the impact of frequency support provision by DFIGs on low-frequency oscillations.
 
 This paper addresses this research gap, by providing new insights on the impact of inertial and primary frequency support provided by DFIGs on other aspects of power system stability, in particular electromechanical and converter-control based low-frequency oscillations. It is shown that, depending on the nature of low-frequency oscillations, inertial and primary frequency support by DFIGs has different impacts.
 First, this impact is demonstrated through the mathematical analysis of the single machine infinite bus (SMIB). Additional simulations on the Kundur two-area system are performed to determine the effect of primary frequency support by DFIGs for different scenarios of wind integration. Furthermore, a sensitivity analysis on droop control parameters characterizing inertial and primary frequency support is carried out to show the impact of parameter selection on low-frequency oscillations.










\section{Power system stability}


\label{sec:small_signal}
Power system stability can be defined as the ability of the system to return to an equilibrium  point after a disturbance. It can be classified according to the observed variables and to the magnitude of the disturbance in rotor angle, voltage and frequency stability. The classification was recently revised to take into account the contribution of converter-interfaced generation, introducing the new concepts of converter-driven and resonance stability \cite{hatziargyriou2020definition}. This paper addresses the frequency, small-signal rotor angle stability and their mutual interactions when DFIGs provide inertial and primary frequency support to the grid.

\subsection{Frequency stability}
Frequency stability refers to the ability of a power system to maintain a steady value of the frequency, following a severe system perturbation. It depends on the ability to sustain or restore equilibrium between system generation and load.

The aggregated second order dynamic model of a power system \eqref{eq:dyn_mod} provides useful insights on frequency stability and the need for frequency support provision \cite{tielens2016relevance}:

\begin{equation}
    2H_{sys}\frac{df_c}{dt}=P_{g}+P_{reg}(f_c)-P_{l}
    \label{eq:dyn_mod}
\end{equation}
where $H_{sys}$ represents the total inertia of the system, $f_c$ the frequency of the center of inertia, $P_g$ the total generation and $P_l$ the load. The term $P_{reg}$ is the regulating power, which is a function of the frequency. 

In steady state, the total generation is equal to the total load. However, when the equilibrium is perturbed, the frequency deviates from the nominal value; this causes the inertial response of synchronous generators, which has a stabilizing effect on the grid. The inertial response is followed by the primary control, whose aim is to bring the system back to an equilibrium point through $P_{reg}(f_c)$. DFIGs do not provide natural inertial response, but appropriate control loops can be implemented in the power converters to provide both inertial and primary frequency support, contributing to the regulating power $P_{reg}(f_c)$.

\subsection{Small-signal stability}
Small-signal stability refers to the transient behavior of power systems, induced by small perturbations around an equilibrium point.

An operating point is considered stable if and only if all eigenvalues of the linearized model of the power system have a negative real part. A damping ratio $\zeta_i$, calculated through \eqref{eq:damp}, can be associated with each complex eigenvalue $\lambda_i=\alpha_i+j\beta_i$, which represents an oscillating mode. 
 
 \begin{equation}
 \label{eq:damp}
 	\zeta_i=\frac{-\alpha_i}{\sqrt{\alpha_i^2+\beta_i^2}}
 \end{equation}
The damping ratio is a useful index to assess how close to instability each oscillating mode is.
 A mode can be defined as critical if $\zeta_i \leq 5 \%$ \cite{oscillations}.

Depending on their frequency, electromechanical oscillating modes in power systems can be distinguished into local, if their frequency is around 2 Hz or inter-area, characterized by a lower frequency from 0.1 to 1 Hz.

As observed in \cite{quintero2014impact}, the integration of converter-interfaced generation introduces oscillating modes whose frequency is similar to both inter-area and local modes, although they depend on control systems and their physical nature is not electromechanical.

Participation factors are used to distinguish electromechanical modes from converter control-based ones.
The participation factor $p_{ik}$, which accounts for the participation of the $k$-th state to the $i$-th mode, can be calculated as in \eqref{eq:part_fact}:
\begin{equation}
    p_{ik}=u_{ki}v_{ki}
    \label{eq:part_fact}
\end{equation}

where $u_{ki}$ and $v_{ki}$ are the elements of the modal matrices of left and right eigenvectors, $\boldsymbol{U}$ and $\boldsymbol{V}$.

Participation factors can be employed to calculate the converter control-based generation participation index (CCBG-PI), introduced in \cite{quintero2014impact}.

The CCBG-PI  of a mode $i$ can be computed from \eqref{eq:CCBG}.

\begin{equation}
    \text{CCBG-PI}_i=\frac{\sum_k^{CCBG}p_{ki}}{\sum_k^{all} p_{ki}} \quad \text{for mode } i 
\label{eq:CCBG}
\end{equation}

This index is employed to identify to what extent the dynamics of a converter-interfaced generator impacts the existing modes and to distinguish converter control-based oscillating modes from inter-area and local ones.

\section{Frequency support provision by DFIG}

During normal operation, DFIG controllers aim at maintaining the turbine at its optimal speed, associated with the maximum power extraction from wind \cite{slootweg2003general}.



\begin{figure}[ht]
\centering
\includegraphics[width=0.5\linewidth]{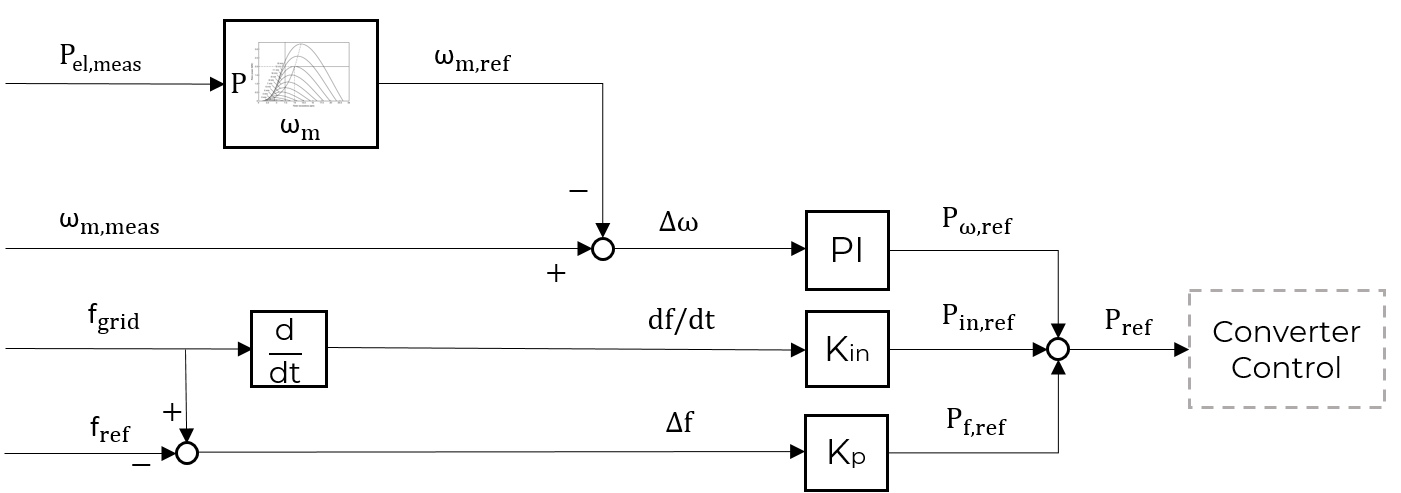}
\caption{Inertial and primary frequency support control scheme \cite{morren2006wind}.}
\label{fig:f_supp_control}
\end{figure}

The maximum power is achieved by controlling the rotor speed $\omega_r$ and the pitching angle $\beta$ to maintain the aero-dynamic power efficiency $C_p$ at its optimal value. This control scheme is called Maximum Power Point Tracking (MPPT) and is represented by the first block of Fig. \ref{fig:f_supp_control}, the PI control.

The reference rotor speed $\omega_{m,ref}$ is estimated using the characteristic curve of the wind turbine and it is used to set the optimal value of the electrical torque reference $T_{\omega,ref}$.

This control system can be enhanced for emulated inertia and primary frequency control, which are represented by the droop controls in the second and third blocks of Fig. \ref{fig:f_supp_control}, respectively. The inertial and primary frequency droop controls, based on ROCOF, $\frac{df}{dt}$, and frequency deviation, $\Delta f$, add two set points to the optimal point $P_{opt}$ resulting from the MPPT. The inertial control is proportional to the controller constant $K_{in}$, the primary frequency control is proportional to $K_p$ and according to \cite{van2015droop}, they can be optimally tuned. The complete control scheme, shown in Fig. \ref{fig:f_supp_control}, is illustrated numerically by~\eqref{eq:opt_control}. The sum of MPPT, primary and inertial droop control sets the reference power $P_{ref}$, required by converter controls.

\begin{equation}
    P_{ref}=P_{opt}-K_{p}\Delta f -K_{in} \frac{df}{dt}
    \label{eq:opt_control}
\end{equation}

In order to work properly, the system requires extra energy  for regulation. For inertial response, since the control has to be fast, a possible option is extracting the kinetic energy from the rotor \cite{keung2008kinetic}.

Since wind turbines rotate to convert wind energy, they store a certain amount of it as kinetic energy. Through specific controls, the kinetic energy can be extracted and provided to the system when needed. Available methods to supply inertial and primary frequency support include, among others, kinetic energy extraction, pitch de-loading or accelerative de-loading \cite{attya2019novel}.

Unlike fuel, which is assumed to be always available, wind has an intrinsically stochastic nature, which makes reserve dispatching difficult. This challenge could be managed by associating energy storage systems to the wind farm providing frequency support, so that the service is available even when the wind is not \cite{choi2016hybrid}.





\section{Mathematical insights from SMIB system}
\label{sec:SMIB}
The SMIB system is commonly employed in power system stability studies to gain useful mathematical insights about power systems \cite{kundur2007power}. In this paper it is used to evaluate mathematically the effect of the inertial and primary frequency control provided by a wind farm on the oscillating modes characterizing the system.

Suppose to add a wind farm with the droop control described in Fig. \ref{fig:f_supp_control} to the generating bus of the SMIB system. The corresponding linearized equations of the system are shown in \eqref{eq:SMIB}:
\begin{equation}
\label{eq:SMIB}
\begin{aligned}
       &2H\dot{\Delta f}=-K_p\Delta f - K_{in} \dot{\Delta f} -K_S\Delta \delta- K_D\Delta f \\
        &\dot{\Delta \delta}=\omega_0\Delta f
\end{aligned}
\end{equation}

where $H$ is the inertia constant of the synchronous generator, $K_S$ the synchronizing coefficient $K_D$ the damping coefficient, $\omega_0=2\pi f_0$ the nominal rotor speed and ${\Delta f}$, ${\Delta \delta}$ the p.u. frequency and rotor angle deviations.

Written in matrix form, the equations \eqref{eq:SMIB} become:
\begin{equation}
    \begin{bmatrix}
    \dot{\Delta f} \\ \dot{\Delta \delta}
    \end{bmatrix}= \begin{bmatrix}
    \frac{-K_p-K_D}{2H+K_{in}} &\frac{-K_S}{2H+K_{in}} \\
    \omega_0 & 0
    \end{bmatrix}\begin{bmatrix}
    \Delta f \\ \Delta \delta
    \end{bmatrix}
\end{equation}

\begin{figure}[ht]
\centering
\includegraphics[width=0.5\linewidth]{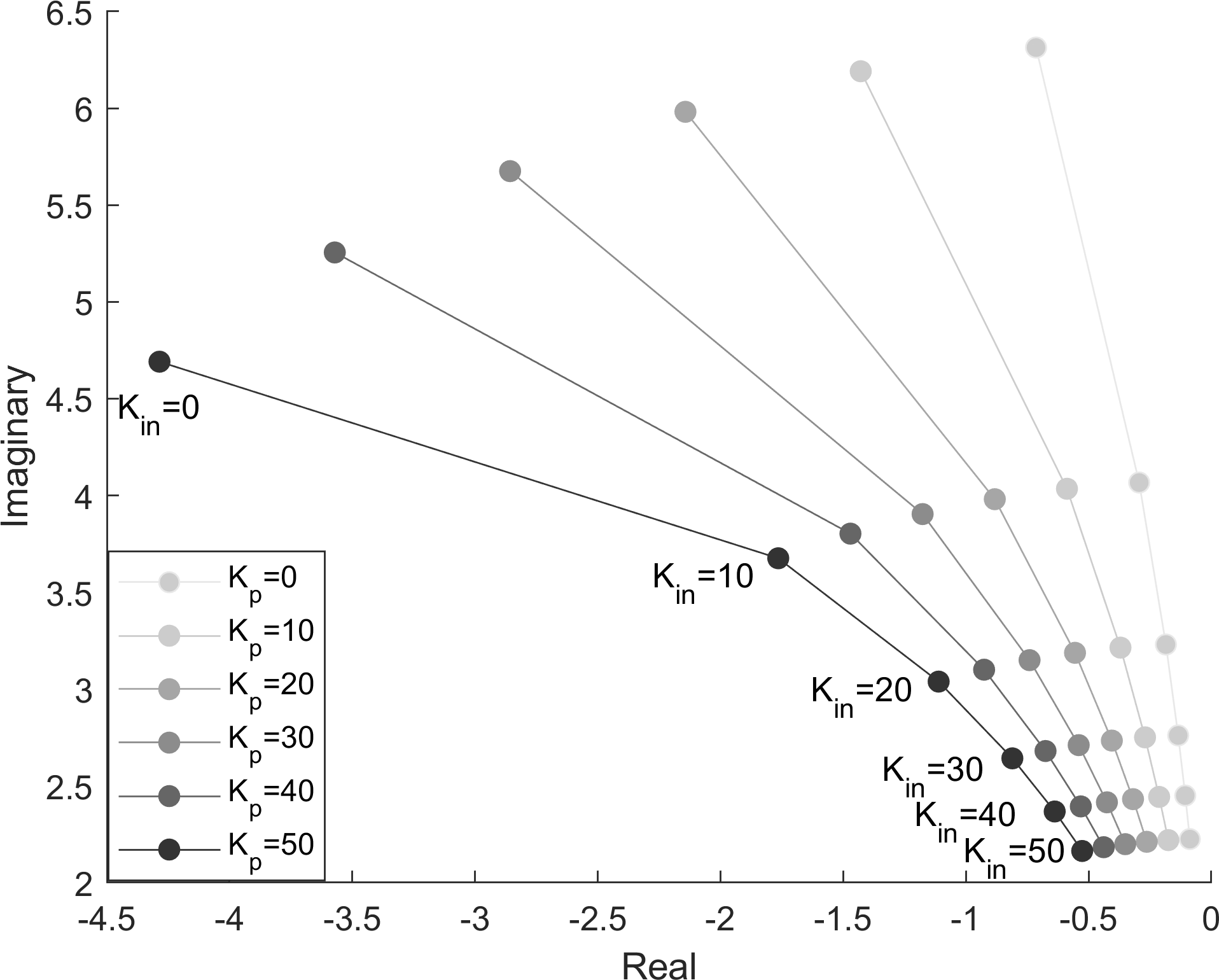}
\caption{Sensitivity analysis of $\lambda_{SMIB}$ to $K_p$ and $K_{in}$.}
\label{fig:SMIB_sens}
\end{figure}

The complex conjugate pair of eigenvalues $\lambda_{SMIB}$ corresponding to this system is equal to:
\begin{equation}
\begin{split}
    \lambda_{SMIB}=&-\frac{K_D+Kp}{4H+2K_{in}} \pm \\ &\frac{\sqrt{(K_p+K_D)^2-8K_SH\omega_0-4K_{in}K_S\omega_0}}{4H+2K_{in}}
\end{split}
\end{equation}
The resulting damping ratio from \eqref{eq:damp} is equal to:
\begin{equation}
    \zeta = \frac{K_p+K_D}{\sqrt{2(K_p+K_D)^2-8K_SH\omega_0-4K_{in}K_S\omega_0}}
\end{equation}

By substituting the values of $K_D=10$, $K_S=0.75$, $H=3.5$ s, $\omega_0=2\pi60$ rad/s from \cite{kundur2007power}, the value of $\lambda_{SMIB}$ can be plotted for different $K_{in}$ and $K_p$, to evaluate its sensitivity of the oscillating modes to the inertial and primary response. Fig.~\ref{fig:SMIB_sens} shows that the primary response increases the damping of the system, whereas the inertial response may have a negative impact on damping. This result will be confirmed in the case study, when assessing the sensitivity of oscillating modes to the inertial and primary droop constant for the Kundur two-area system, a widely employed benchmark system for stability studies.

\section{Case Study}

In this section, a modified version of the two-area Kundur system \cite{kundur2007power} including wind generation is analyzed with Matlab/Simulink to investigate the effects on small-signal stability of DFIGs providing frequency support to power systems.

\begin{figure}[ht]
\centering
\includegraphics[width=0.7\linewidth]{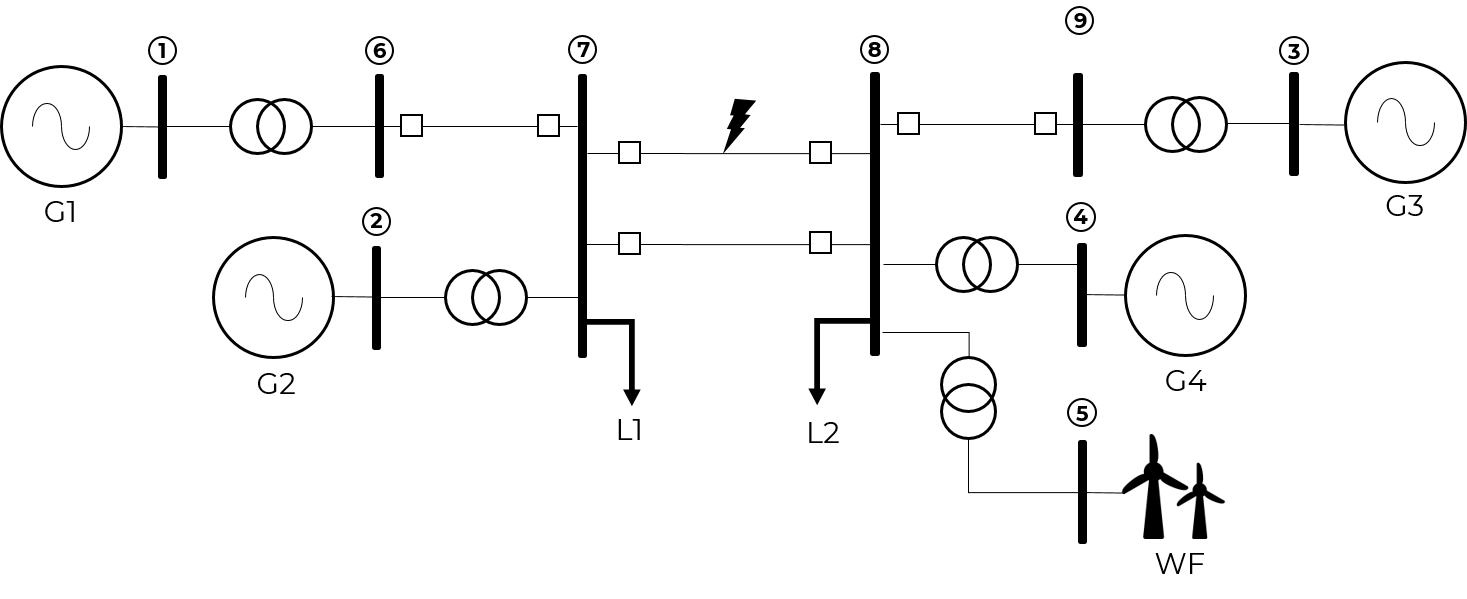}
\caption{Modified two-area Kundur system.}
\label{fig:mod_2_area}
\end{figure}

The one-line diagram of the system is drawn in Fig. \ref{fig:mod_2_area}.

\begin{figure}[ht]
\centering
\includegraphics[width=0.7\linewidth]{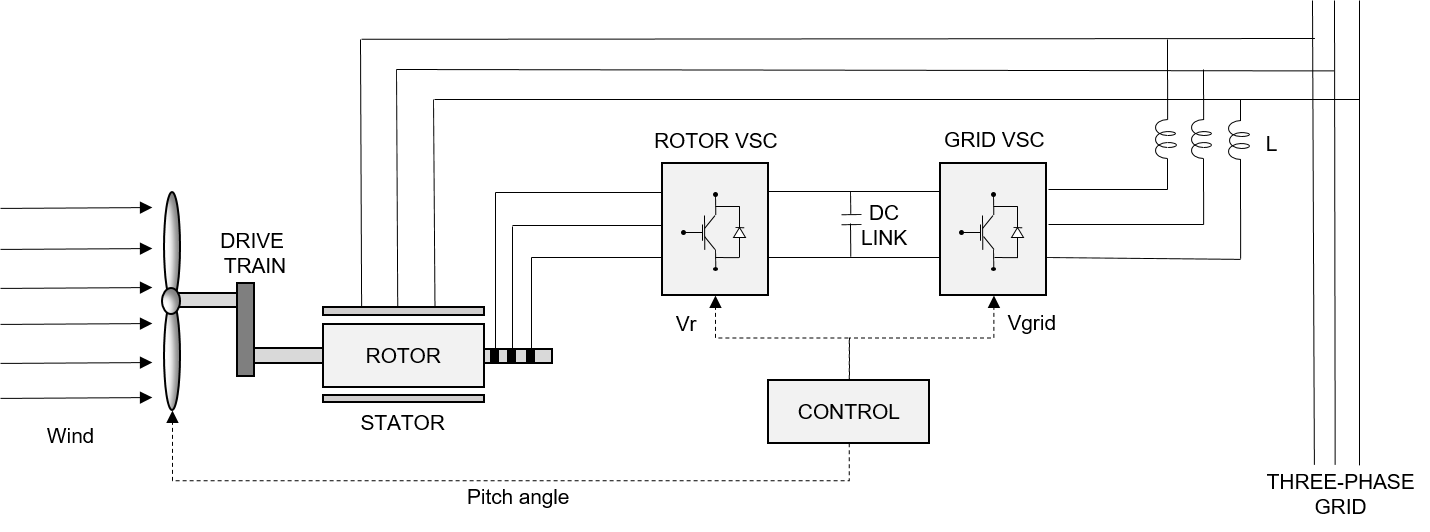}
\caption{Structure of the DFIG wind turbine model.}
\label{fig:DFIG}
\end{figure}

The original system consists in two areas, each having two synchronous generators. The areas are connected by two parallel HV transmission lines of 220 km length, at 230 kV.
All synchronous generators are modelled in detail, including prime movers, speed governors, excitation systems and PSSs.Their nominal power is equal to 900 MVA each and the active power generated in the considered operating condition is around 600 MW. Each area has a single lumped load, modelled as constant impedance and the total load is around 2300 MW. Wind power generation, added at bus 5, is represented by a lumped dynamical model of a wind farm working with DFIG type turbines. Its nominal power depends on the scenario considered. The model of a DFIG is represented in Fig.~\ref{fig:DFIG}. It consists in a simplified mechanical model of the turbine, the asynchronous machine model, rotor and grid-side voltage source converters (VSC) and the related controls. Wind speed over the wind farm is assumed to be constant during the observation time interval and spatially uniform \cite{slootweg2003aggregated}.

For the purpose of this case study, DFIGs are equipped with the inertial and primary frequency control scheme presented in Fig. \ref{fig:f_supp_control}.

A first case study is carried out considering only primary frequency support provision, followed by a sensitivity analysis on the effects of $K_{in}$ and $K_p$ on low-frequency oscillation damping. 


\subsection{Impact of primary frequency support provision by DFIGs}

After a critical review of the state of art, the authors of \cite{du2017small} suggested that a complete characterization of the impacts of DFIGs in power system’s small-signal stability analysis should include the following scenarios: 

\begin{itemize}
    \item Scenario A: initial scenario without the wind farm, corresponding to the classical Kundur two-area system.
    \item Scenario B: a DFIG-based wind farm is added to the system, at bus 5, along with synchronous generator $\text{G}_4$. In this case, the nominal power is chosen to be 300 MVA.
    \item Scenario C: the wind farm displaces synchronous generator $\text{G}_4$ and its nominal power is chosen to be around 600 MVA, in order to supply the power lacking from $\text{G}_4$.
\end{itemize}

In this paper, for each scenario involving the presence of the wind farm, the following sub-scenarios are explored:
\begin{itemize}
    \item Voltage or reactive power control mode;
    \item With or without primary frequency support provision.
\end{itemize}

All the combinations are examined, for a total of nine cases, including scenario A.

\subsubsection{Scenario A (no wind)}

the analysis of the original two-area Kundur system, without the inclusion of the wind farm, provides an initial understanding of the transient behavior of the system and a foundation to assess the impact of wind farms providing frequency support on low-frequency oscillations. 
The system parameters are tuned to obtain a loosely damped starting system, which results in a more easily observable damping effect of the wind farm. This is achieved by setting the proportional gain of PSSs to $K_{pss}=10$.

We start by investigating the small-signal stability of this system. Tables \ref{tab:local_modes_poor_damp} and \ref{tab:interarea_modes_poor_damp} present four local $\lambda_{l,i}$ and inter-area $\lambda_{ia,i}$ modes, which are the most critical oscillation modes, i.e. the ones closer to the imaginary axis and with the lowest damping ratio.

\setlength\doublerulesep{0.5pt}
\begin{table}[ht]
\centering
\caption{Local modes (Scenario A)}
\label{tab:local_modes_poor_damp}
\resizebox{.5\linewidth}{!}{
\begin{tabular}{cccc}
\toprule[1pt]
\toprule[0.3pt]
    & Eigenvalue             & Damping & Frequency [Hz] \\ \midrule
$\lambda_{l,1}$ & $-20.597 \pm18.900i $ & $0.736$   &   $3.008$ \\ 
$\lambda_{l,2}$ & $ -18.232 \pm14.251i $ & $0.787$   & $2.268$ \\ 
$\lambda_{l,3}$ & $ -1.026 \pm 8.566i $ & $\textbf{0.119}$   & $1.363$ \\ 
$\lambda_{l,4}$ & $-2.048 \pm 6.578i $ & $0.297$   & $1.047$ \\ 
\bottomrule[0.3pt]
\bottomrule[1pt]
\end{tabular}
}
\end{table}

\begin{table}[ht]
\centering
\caption{Inter-area modes (Scenario A)}
\label{tab:interarea_modes_poor_damp}
\resizebox{.5\linewidth}{!}{
\begin{tabular}{cccc}
\toprule[1pt]
\toprule[0.3pt]
    & Eigenvalue             & Damping & Frequency [Hz] \\ 
\midrule
$\lambda_{ia,1}$ & $ -38.981 \pm 1.220i $ & $0.999$   & $0.194$\\ 
$\lambda_{ia,2}$ & $ -25.819 \pm 0.915i $ & $0.999$   & $0.145$ \\ 
$\lambda_{ia,3}$ & $-24.498 \pm 0.801i$  & $0.999$   & $0.127$ \\ 
$\lambda_{ia,4}$ & $-0.046 \pm 2.877i $ & $\textbf{0.016}$   & $0.458$ \\ \bottomrule[0.3pt]
\bottomrule[1pt]
\end{tabular}
}
\end{table}

Inter-area modes appear well damped, excluding $\lambda_{ia,4}$, a critical mode with a very low damping ratio $\zeta_{ia,4}=1.6\%$. Local modes are well damped too, with $\lambda_{l,3}$ showing the lowest damping ratio $\zeta_{l,3}=11.9\%$. These two modes are the dominant ones; hence, the following scenarios including wind integration will focus on them.

\begin{figure}[ht]
\centering
\includegraphics[width=0.5\linewidth]{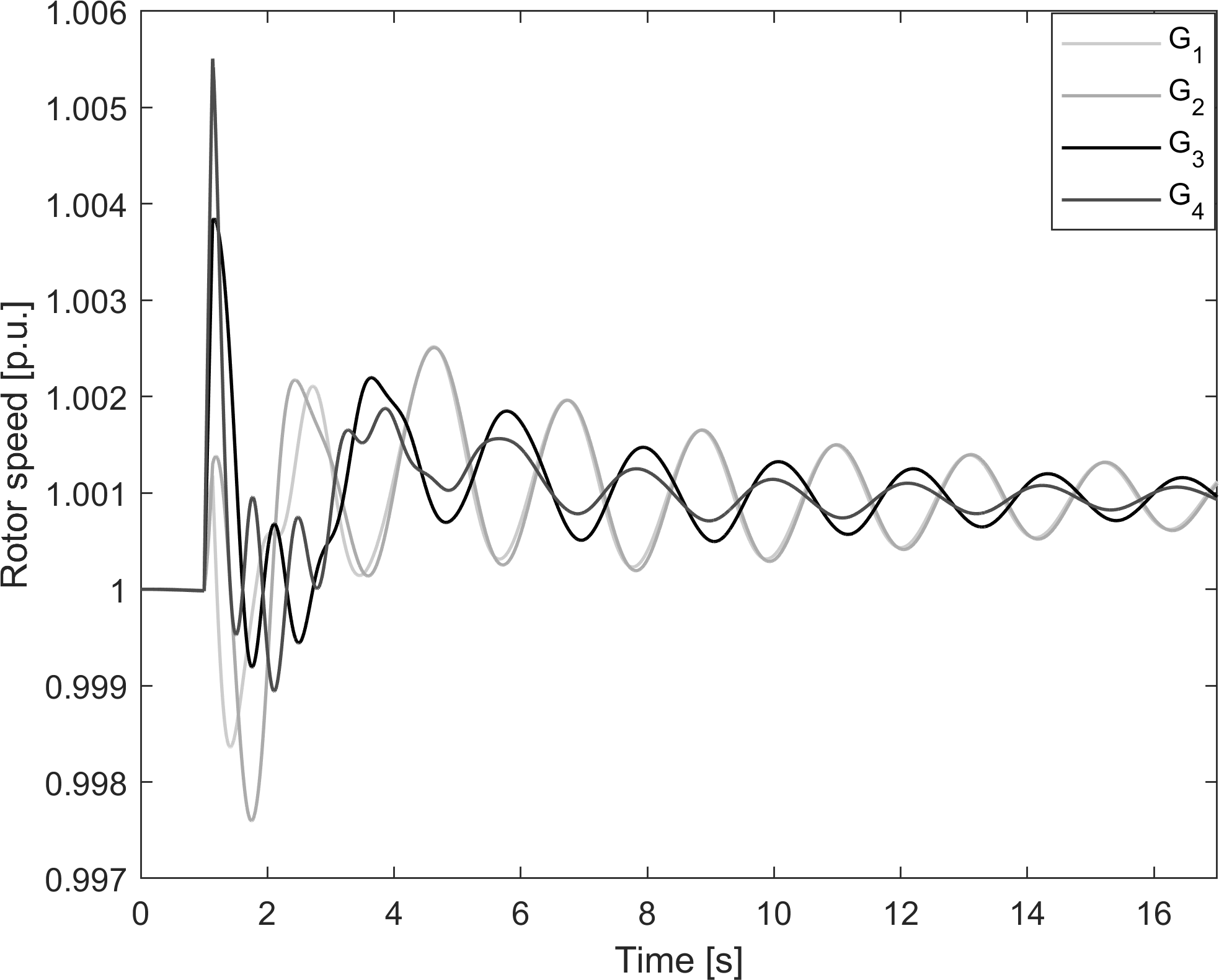}
\caption{Time domain simulation considering a three-phase fault on the tie-line.}
\label{fig:tds_no_wind}
\end{figure}

The results of small-signal stability analysis are complemented with a time domain simulation, considering a three-phase fault lasting ten cycles on the tie-line connecting Area 1 and Area 2. The result, in terms of rotor speeds of synchronous generators, is shown in Fig. \ref{fig:tds_no_wind}. The figure shows the generators of Area 1 oscillating against the ones of Area 2, with a frequency around $0.5$  Hz, as predicted by the small-signal stability analysis.

\begin{figure}[ht]
\centering
\includegraphics[width=0.5\linewidth]{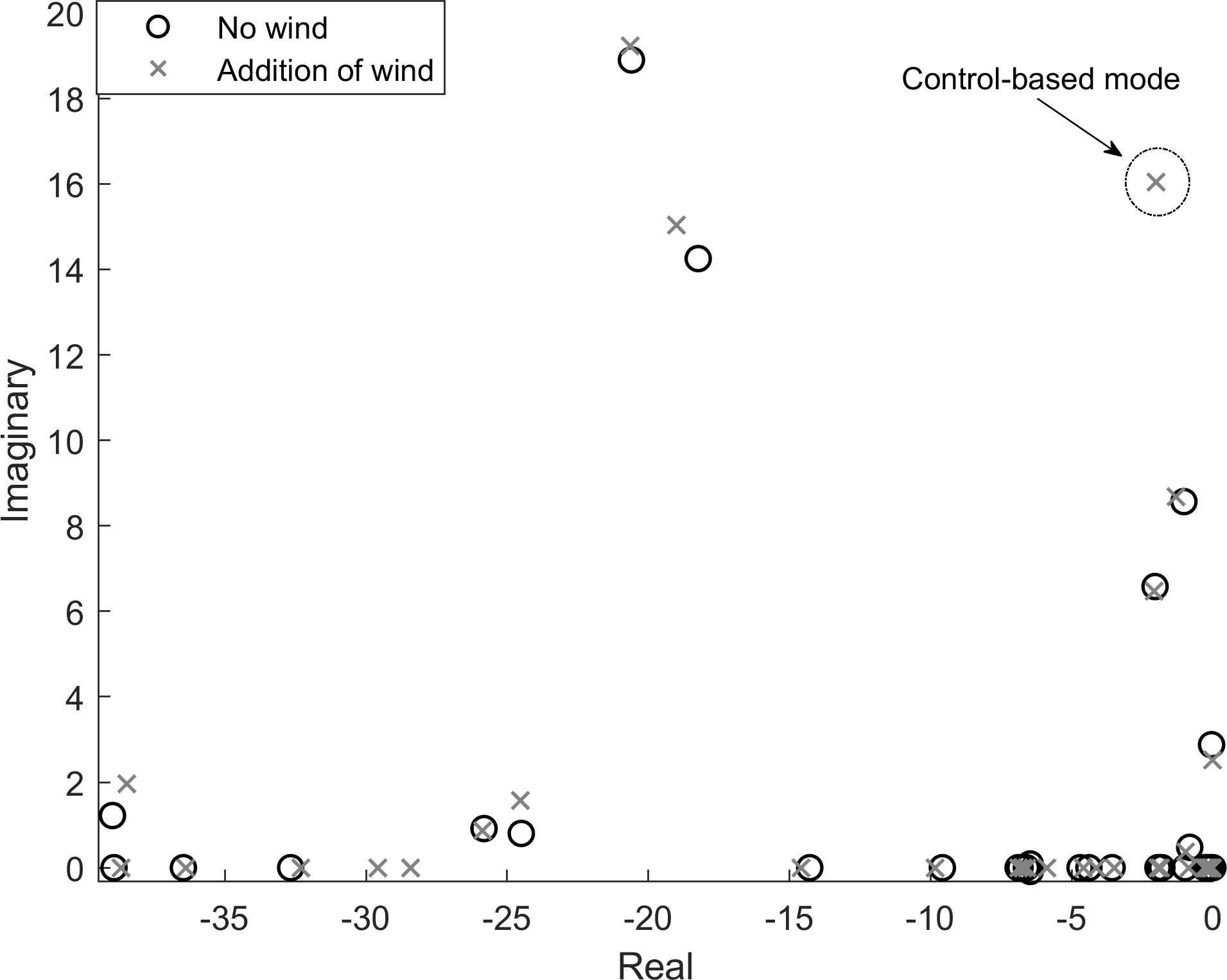}
\caption{Comparison of eigenvalues between Scenario A and B.}
\label{fig:comparison_wind}
\end{figure}

\subsubsection{Scenario B (addition of the wind farm)}

the addition of the wind farm leads to a change of power flows and to dynamic interactions introduced by the controls of DFIGs. These joint effects lead to the movement of both local and inter-area modes. Fig. \ref{fig:comparison_wind} illustrates how oscillating modes change as the wind farm is added to the system.

\begin{figure}[ht]
    \centering
    \includegraphics[width=0.5\linewidth]{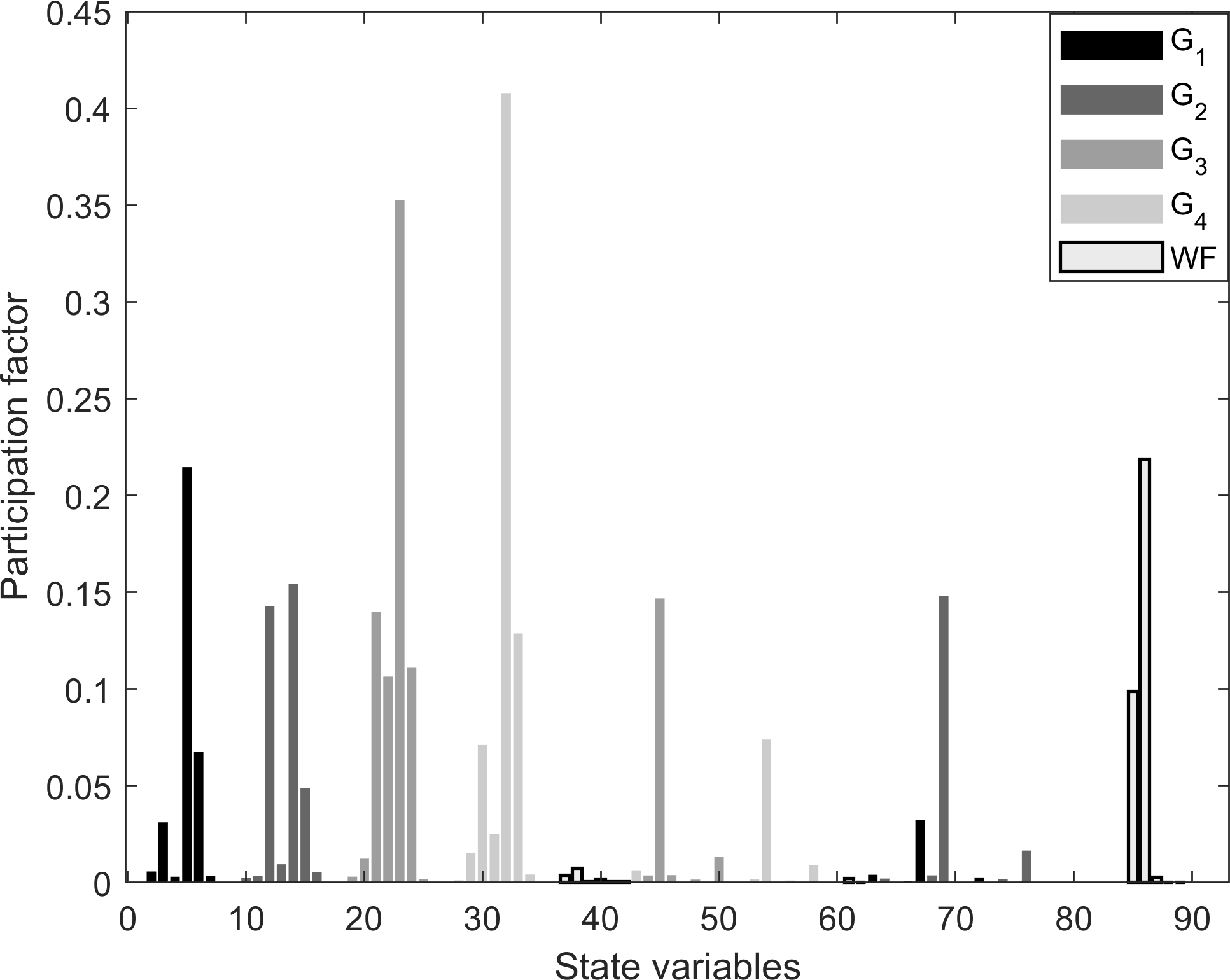}
    \caption{Participation factors associated with the inter-area mode $\lambda_{ia,3}$.}
    \label{fig:autoval_29}
\end{figure}

The dynamic interaction of the wind farm is observable through the analysis of participation factors.
Fig. \ref{fig:autoval_29} shows the participation factors of inter-area mode $\lambda_{ia,3}$. In addition to the participation of all synchronous generators, given its inter-area nature, a non-negligible participation of the wind farm is observed, represented by the non-zero values of the participation factors associated with the wind farm state variables. This leads to a non-zero, although little, CCBG-PI for $\lambda_{ia,3}$.

An important impact is the emergence of a new oscillating mode, in the range of local modes, which is a converter control-based mode, defined as $\lambda_{cc}$. This mode is circled in Fig. \ref{fig:comparison_wind}.  Its nature can be distinguished from electromechanical modes by the analysis of participation factors and the calculation of the CCBG-PI with \eqref{eq:CCBG}. The CCBG-PI of $\lambda_{cc}$ is equal to $98.6 \%$, denoting a negligible participation of synchronous generators to this mode. Hence, $\lambda_{cc}$ is a converter-control based mode. 
 
Table \ref{tab:summary} illustrates the numerical values of the dominant inter-area, local and converter control-based modes for the nine cases of interest.

The addition of the wind farm affects the dominant inter-area mode, whose damping ratio decreases from $\zeta_{ia}=3.30\%$, corresponding to the case without wind, to $\zeta_{ia}=0.34\%$ when the wind farm is added and voltage control is employed. The mode crosses the imaginary axis when reactive power control is employed, resulting in the positive real component of the dominant inter-area mode, highlighted in bold in Table \ref{tab:summary}.

The dominant local mode, on the other hand, improves slightly from $\zeta_{l}=12.8\%$ of the scenario without wind to around $\zeta_{l}=15\%$, when the wind farm is added.

The converter control-based mode seems well damped, with $\zeta_{cc}=12.5\%$ and does not seem to be affected by the control mode of the DFIGs.

At this point, the primary frequency support provision is added to the wind farm and its effect on the dominant modes is investigated.

For this scenario, adding the primary frequency support capability increases the small-signal stability of the system. The damping ratio of the inter-area mode increases from $\zeta_{ia}=0.34\%$, without frequency support, to $\zeta_{ia}=2.80\%$.
This value is still lower than the one in the scenario without the wind. However, the frequency support capabilities show a beneficial effect on the dominant inter-area mode.

Regarding the dominant local mode, the provision of frequency support by the wind farm additionally increases its damping ratio from $\zeta_l=12.8\%$ in the case without wind to around $\zeta_l=20.3\%$ when the wind farm providing primary frequency support is added.

An interesting result is associated with the dominant converter control-based mode. Indeed, frequency support appears to negatively affect its damping ratio, which decreases from $\zeta_{cc}=12.5\%$, without frequency support, to $\zeta_{cc}=1.90\%$ with frequency support. Furthermore, a decrease of the CCBG-PI when adding the frequency support was observed. This indicates a larger participation of synchronous generators in the converter control-based mode, which may stem from DFIGs' active power responding to the system's frequency deviations. 

The effect of employing voltage or reactive power control appears to be negligible, around $\pm 1\%$ of the damping ratios for all the dominant modes.

\begin{figure}[ht]
    \centering
    \includegraphics[width=0.5\linewidth]{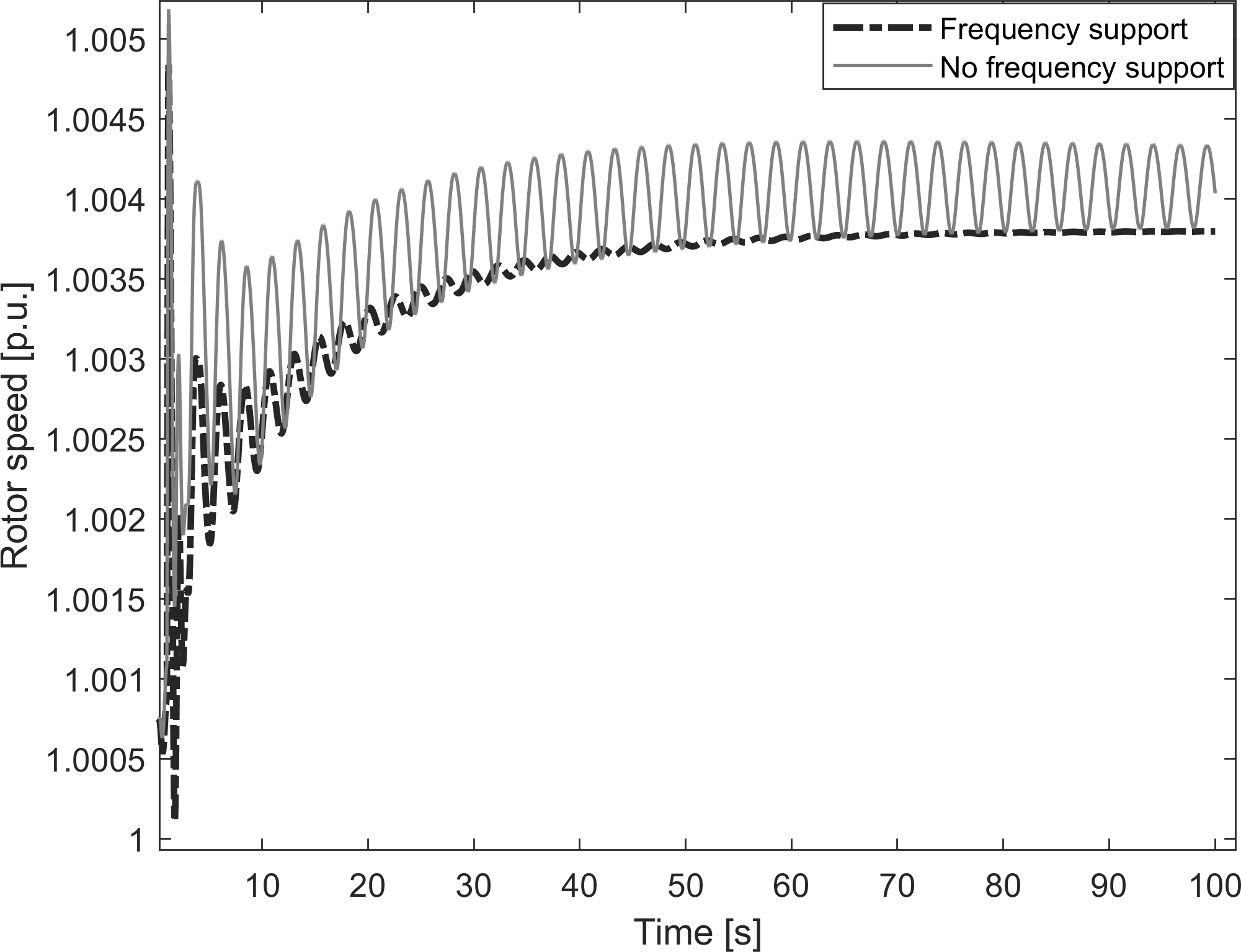}
    \caption{Rotor speed of generator $\text{G}_3$ with and without primary frequency support by the wind farm.}
    \label{fig:primary_TDS}
\end{figure}

The time domain simulation, considering a three-phase fault on the tie-line as in the previous scenario, shows the impact of frequency support provision on the oscillations. Fig. \ref{fig:primary_TDS} illustrates the rotor speed of synchronous generator $\text{G}_3$ when wind is added and allows a comparison of the cases with and without primary frequency support. 

As expected from the eigenvalue analysis, since the dominant inter-area mode is close to a Hopf bifurcation, i.e. it is crossing the imaginary axis, the system suffers sustained oscillations in the scenario with the wind farm without frequency support capabilities. On the other hand, when the wind farm provides primary frequency support, the amplitude of the oscillations is reduced and they are damped out after a brief time interval. The other synchronous generators, not illustrated, show a similar behavior.

\begin{figure}[ht]
    \centering
    \includegraphics[width=0.5\linewidth]{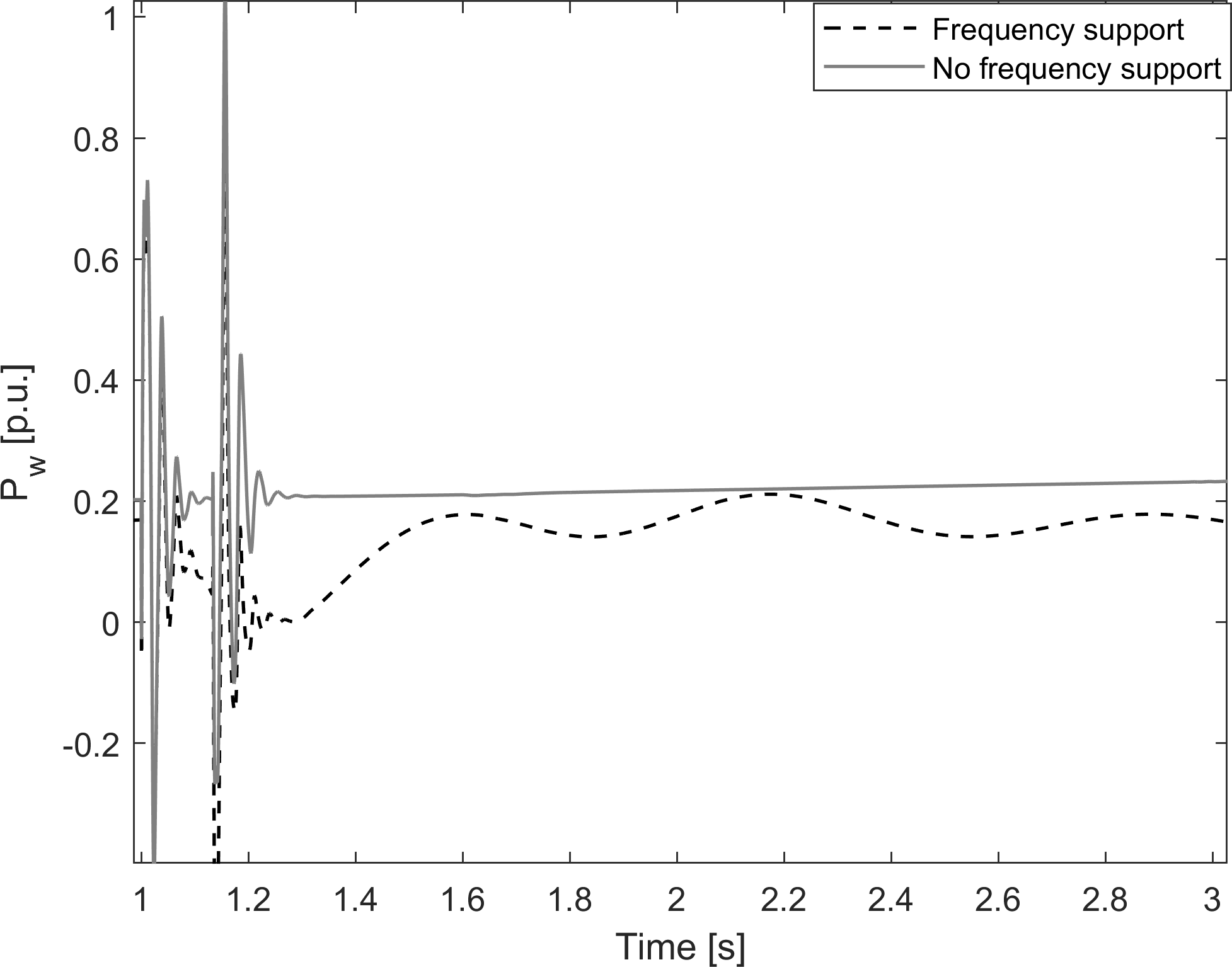}
    \caption{Active power from the wind farm with and without primary frequency support.}
    \label{fig:P_freq_supp}
\end{figure}

Primary frequency support provision by the wind farm requires the DFIGs to change their operating point from the optimal one in response to a frequency deviation. Fig.~\ref{fig:P_freq_supp} illustrates the active power provided by the DFIGs with and without frequency support, during the first seconds after the occurrence of the fault. The DFIGs would normally be insensitive to the frequency deviations of the system. However, the droop control allows them to respond to these deviations, improving the frequency stability of the system.

\begin{table*}[ht]
\caption{Result summary for all scenarios considering dominant inter-area, local and converter control-based modes}
\centering
\resizebox{0.95\linewidth}{!}{
\begin{tabular}{cccccccccc}

\toprule[1pt]
\toprule[0.3pt]
\multicolumn{1}{c}{\multirow{2}{*}{Scenario}}                                                   &       \multicolumn{1}{c}{\multirow{2}{*}{Freq. supp.}}                                             & \multicolumn{1}{c}{\multirow{2}{*}{Control}}   & \multicolumn{2}{c}{Inter-area mode}              & \multicolumn{2}{c}{Local mode}                  & \multicolumn{3}{c}{Converter control-based mode}                                                  \\ \cline{4-10} 
                                                   &                                                 &   & {Eigenvalue}        & Damping & {Eigenvalue}        & Damping & {Eigenvalue}        & {Damping} & {CCBG-PI} \\
                                                   \midrule[0.3pt]
                                                   \midrule[1pt]
                                
{(A) No wind}                      & {-}                          & - & $-0.096 \pm 2.942i$ & $0.033$  & $-1.131\pm 8.757i$  & $0.128$  & {-}                 & {-}       & {-}        \\ 
\midrule[1.3pt]
\multicolumn{1}{c}{\multirow{4}{*}{(B) Add. wind}}   & \multicolumn{1}{c}{\multirow{2}{*}{No f. supp.}} & V & \multicolumn{1}{c}{$-0.008 \pm 2.512i$}  & $0.003$  & \multicolumn{1}{c}{$-1.289 \pm 8.652i$} & $0.147$  & \multicolumn{1}{c}{$-2.036 \pm 16.093i$} & \multicolumn{1}{c}{$0.125$}  & $0.985$  \\ 
\multicolumn{1}{c}{}                             & \multicolumn{1}{c}{}                           & Q & \multicolumn{1}{c}{\textbf{0.026 $\pm$ 2.342i}}  & \textbf{-0.011} & \multicolumn{1}{c}{$-1.345 \pm 8.601i$} & $0.154$  & \multicolumn{1}{c}{$-2.013 \pm 15.942i$} & \multicolumn{1}{c}{$0.125$}  & $0.985$  \\ \cmidrule[0.6pt]{2-10}
\multicolumn{1}{c}{}                             & \multicolumn{1}{c}{\multirow{2}{*}{F. supp.}}    & V & \multicolumn{1}{c}{$-0.069 \pm 2.521i$} & $0.027$  & \multicolumn{1}{c}{$-1.709 \pm 8.233i$} & $0.203$  & \multicolumn{1}{c}{$-0.343 \pm 17.204i$} & \multicolumn{1}{c}{$0.019$}  & $0.726$  \\ 
\multicolumn{1}{c}{}                             & \multicolumn{1}{c}{}                           & Q & \multicolumn{1}{c}{$-0.069 \pm 2.500i$} & $0.028$  & \multicolumn{1}{c}{$-1.724 \pm 8.199i$} & $0.206$  & \multicolumn{1}{c}{$-0.328 \pm 17.108i$} & \multicolumn{1}{c}{$0.019$}  & $0.723 $ \\ \midrule[1.3pt]
\multicolumn{1}{c}{\multirow{4}{*}{(C) Displ. $\text{G}_4$}} & \multicolumn{1}{c}{\multirow{2}{*}{No f. supp.}} & V & \multicolumn{1}{c}{$-0.459 \pm 3.756i$} & $0.121$  & \multicolumn{1}{c}{$-1.671 \pm 7.585i$} & $0.215$  & \multicolumn{1}{c}{$-2.066 \pm 16.009i$} & \multicolumn{1}{c}{$0.128$}   & $0.966$  \\ 
\multicolumn{1}{c}{}                             & \multicolumn{1}{c}{}                           & Q & \multicolumn{1}{c}{$-0.484 \pm 3.731i$} & $0.129$  & \multicolumn{1}{c}{$-1.616 \pm 7.589i$} & $0.208$  & \multicolumn{1}{c}{$-2.218 \pm 16.896i$} & \multicolumn{1}{c}{$0.130$}  & $0.963$  \\ \cmidrule[0.6pt]{2-10}
\multicolumn{1}{c}{}                             & \multicolumn{1}{c}{\multirow{2}{*}{F. supp.}}    & V & \multicolumn{1}{c}{$-0.694 \pm 2.837i$} & $0.238$  & \multicolumn{1}{c}{$-1.658 \pm 7.577i$} & $0.214$  & \multicolumn{1}{c}{$-0.649 \pm 17.696i$}  & \multicolumn{1}{c}{$0.037$} & $0.708$  \\ 
\multicolumn{1}{c}{}                             & \multicolumn{1}{c}{}                           & Q & \multicolumn{1}{c}{{\color{darkgray}\textbf{-0.740 $\pm$ 2.744i}}} & {\color{darkgray}\textbf{0.260}}  & \multicolumn{1}{c}{$-1.611 \pm 7.579i$} & $0.208$  & \multicolumn{1}{c}{$-0.529 \pm 18.245i$}  & \multicolumn{1}{c}{$0.029$}  & $0.722$  \\
\bottomrule[0.3pt]
\bottomrule[1pt]
\end{tabular}
}
\label{tab:summary}
\end{table*}

\subsubsection{Scenario C (displacement of $\text{G}_4$)}

as pointed out by the authors of \cite{du2017small}, a full characterization of the impacts of a wind farm should include both the addition of the wind farm and the displacement of part of the  synchronous generation. 

In this scenario, $\text{G}_4$ is displaced by the wind farm, as it is the closest synchronous generator to the wind farm location. In this way, the effect of wind integration on power flows is minimized.

Table \ref{tab:summary} shows that the addition of a wind farm, with the contextual displacement of a synchronous generator, leads to a great improvement of both local and inter-area mode damping. The damping ratio of the inter-area mode increases from $\zeta_{ia}=3.30\%$ of the base case, to about $12\%$ when adding the wind farm and displacing $\text{G}_4$. The damping of the dominant local mode is almost doubled, from $\zeta_l=12.8\%$ to $\zeta_l=21.5\%$. The effect on the converter control-based mode is similar to the previous scenario.

In this case, probably due to the missing primary frequency control of $\text{G}_4$, the primary frequency support provision by the wind farm has a large beneficial effect on the damping ratio of the inter-area mode. Indeed, this scenario is the best damped one: the damping ratio of the inter-area mode changes from the starting scenario without wind $\zeta_{ia}=3.30\%$ to $\zeta_{ia}=26.0\%$, as shown in Table \ref{tab:summary} (dark-grey bold values).

\begin{figure}[ht]
    \centering
    \includegraphics[width=0.5\linewidth]{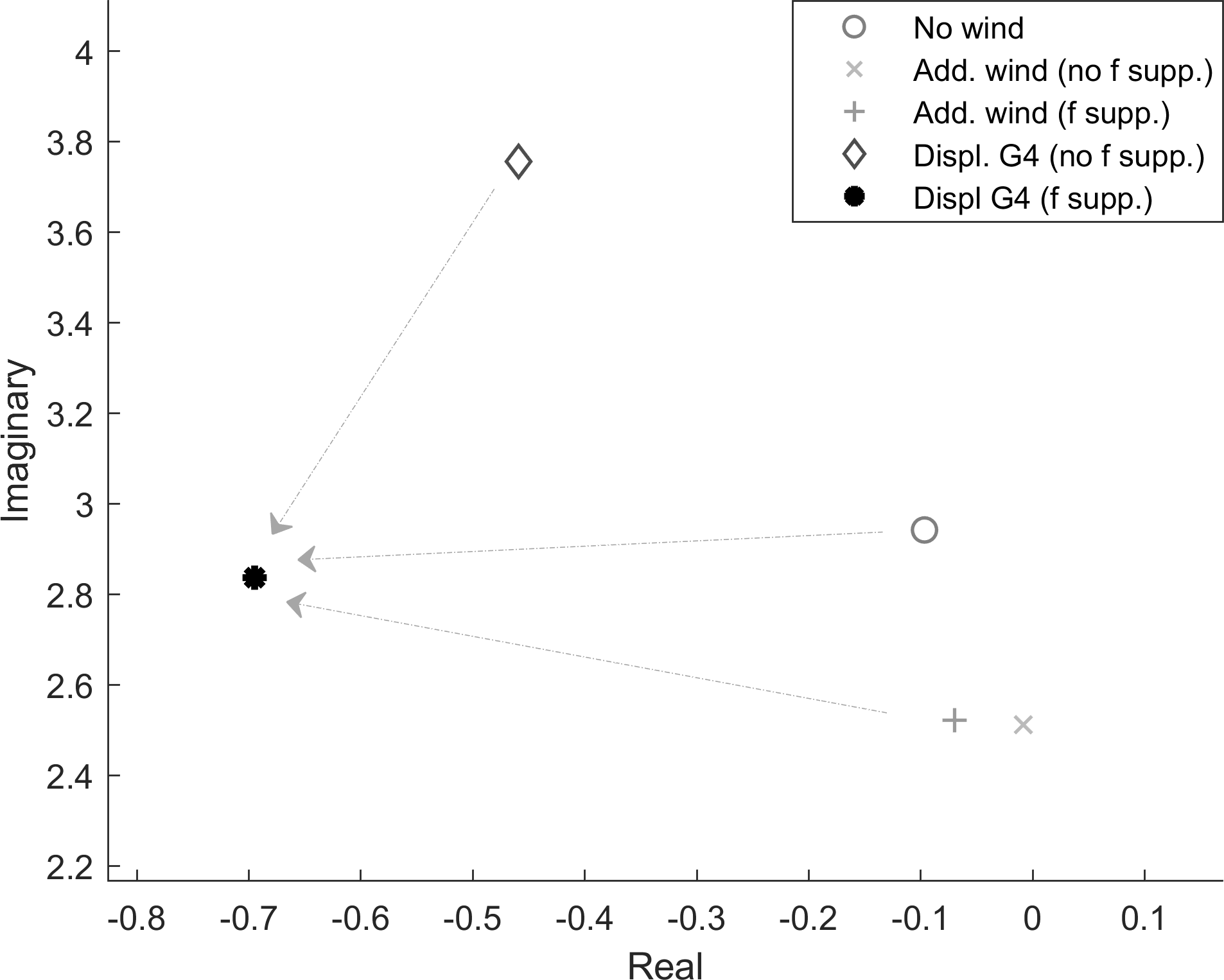}
    \caption{Graphical visualization of inter-area modes for different scenarios.}
    \label{fig:interarea_eigen}
\end{figure}

The change in damping can be visualized by zooming the complex plane around the region of inter-area modes. Fig.~ \ref{fig:interarea_eigen} illustrates the movement of the dominant inter-area mode towards a more stable region.

Although the primary frequency support provision improves the damping of electromechanical oscillations, it reduces the damping of the dominant converter control-based mode. However, since this mode is not an electromechanical one, it may not largely affect the stability of the system as a whole. A time domain simulation is carried out to check the stability of the system, given the negative damping of the converter control-based dominant mode.

\begin{figure}[ht]
    \centering
    \includegraphics[width=0.5\linewidth]{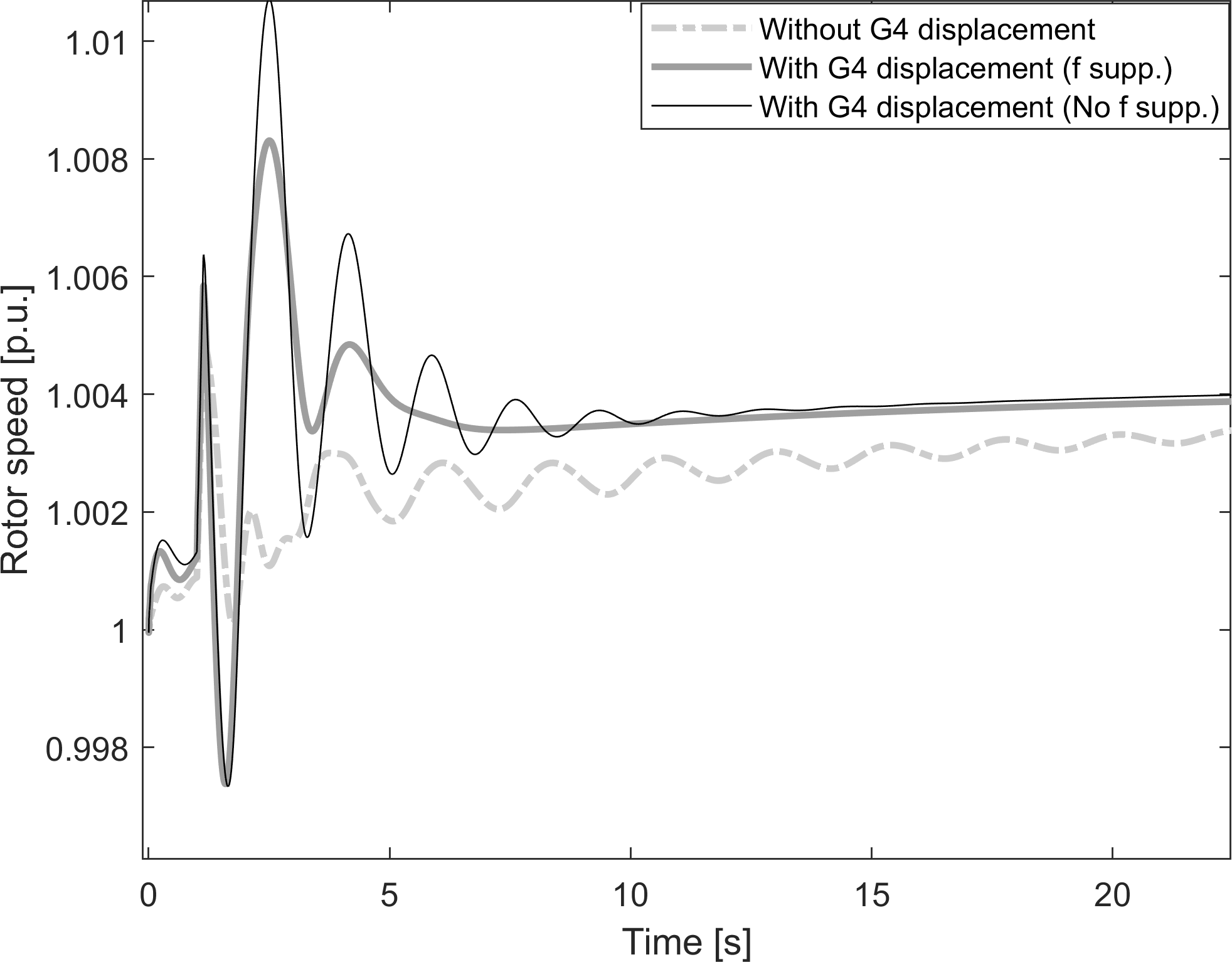}
    \caption{Rotor speed of generator $\text{G}_3$ with wind farm providing primary frequency response, with and without displacement of $\text{G}_4$.}
    \label{fig:primary_TDS_compare}
\end{figure}

In Fig. \ref{fig:primary_TDS_compare}, the rotor speed of $\text{G}_3$ is compared for different scenarios. The one involving $\text{G}_4$ displacement leads to larger oscillations, due to the lacking inertia of the displaced synchronous generator. The figure shows how the primary frequency support results in a reduced amplitude of oscillations in this case. 
Also, Fig. \ref{fig:primary_TDS_compare} shows that the slightly damped control mode does not seem to affect the stability of the system as a whole. However, this interesting link between frequency support and converter controls requires further investigation, which is out of the scope of this paper.

As expected from the previous small-signal stability analysis, in the case with displacement of $\text{G}_4$ and frequency support, having $\zeta_{ia}=26\%$, the oscillations are damped very fast, in about 5 seconds. When no frequency support is provided, $\zeta_{ia}=12.9\%$ and oscillations require about 15 seconds to be damped. If $\text{G}_4$ is not displaced, the frequency support still improves the damping of the system, but to a smaller extent: $\zeta_{ia}=2.80\%$ and the oscillations require more than 20 seconds to be damped.  


\subsubsection{Discussion}

the previous case studies demonstrated that frequency support provision, although implemented mainly to improve the frequency stability after power imbalance, can provide a beneficial effect on the damping of electromechanical oscillations. The effects on small-signal stability can be summarized as follows: 

\begin{enumerate}
    \item The damping of both local and inter-area oscillation modes improves when the wind farm provides primary frequency support. This effect is greater when a wind farm displaces part of the synchronous generation.
    \item The dynamics of the wind farm becomes more intertwined with the inter-area and local oscillating modes. There is a less clear distinction between converter and power system oscillating modes. Indeed the CCBG-PI of $\lambda_{cc}$ is lower, whereas the CCBG-PI associated with local and inter-area modes increases.
    \item The converter control-based mode becomes less damped with the displacement of $\text{G}_4$. This phenomenon does not seem to affect the stability of the overall system, but it requires further investigation.
\end{enumerate}

These results can be generalized to any values of the proportional gain $K_p$. Indeed, the value of $K_p$ only affects the magnitude of the effect. Primary frequency support appears beneficial to inter-area and local modes, detrimental to converter control-based mode and the value of $K_p$ only determines how beneficial or detrimental this effect is. 

On the other hand, the sensitivity of the system to the inertial response provision is less straightforward: different values of $K_{in}$ can lead to either a beneficial or detrimental effect for each mode. The following paragraph provides a sensitivity analysis on both $K_p$ and $K_{in}$.

\subsection{Addition of inertial response and sensitivity analysis on proportional gains $K_p$ and $K_{in}$}

In this case study, the inertial response is included along with primary frequency response and the effect of both on low-frequency oscillations is discussed.

This analysis considers only the case with the addition of the wind farm, working in voltage control mode and without displacement of $\text{G}_4$. The displacement of $\text{G}_4$ leads to similar qualitative results, although the magnitude of the sensitivity changes.

The sensitivity analysis is carried out by calculating  the dominant inter-area, local and control based modes for a set of combinations of $K_p$ and $K_{in}$, which vary in the range between 0 and 50, with a discrete step of 10. The total number of cases considered is 36, enough to draw the sensitivity curves of eigenvalues on the complex plane.

\begin{figure}[ht]
    \centering
    \includegraphics[width=0.5\linewidth]{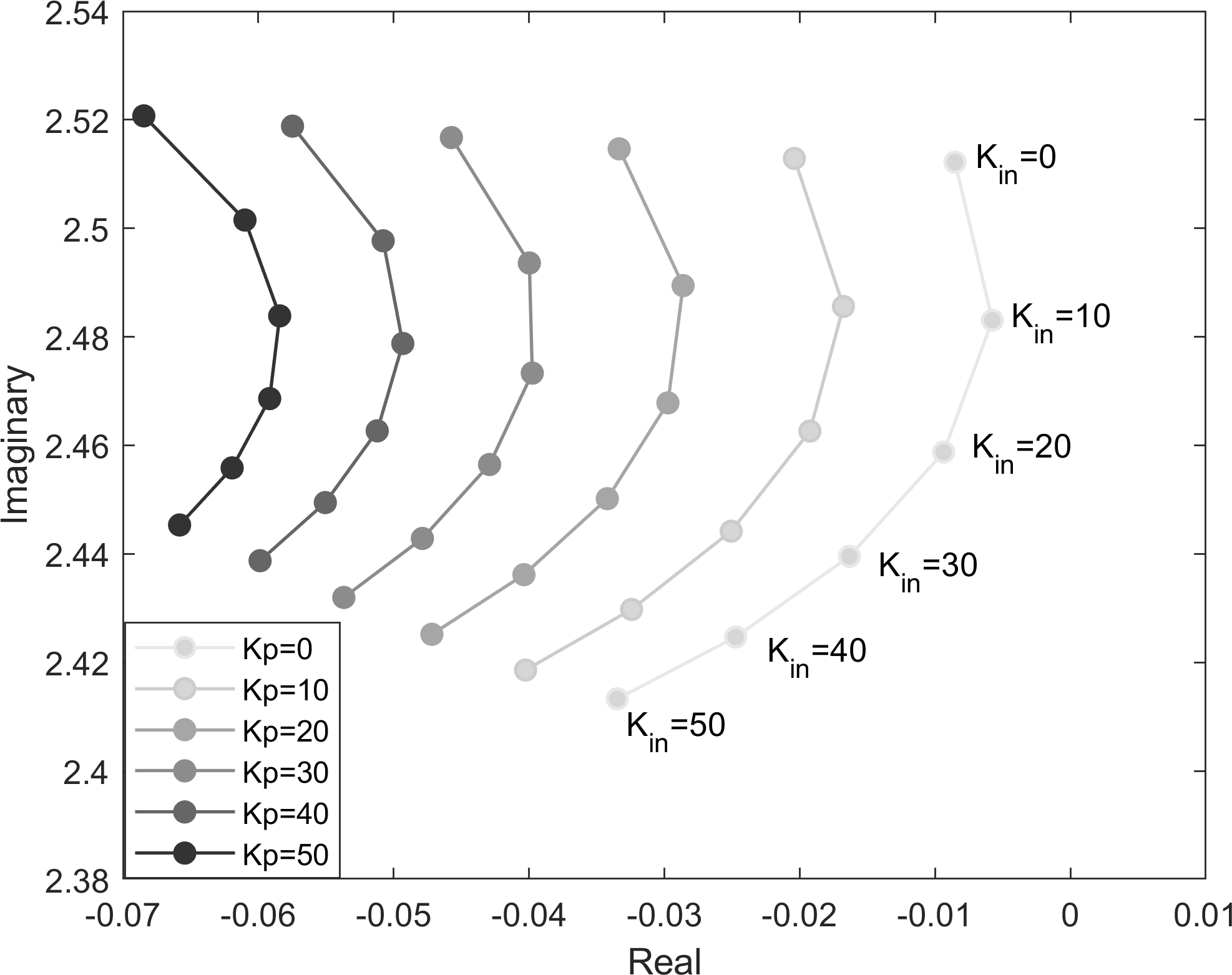}
    \caption{Sensitivity analysis of inter-area modes to $K_p$ and $K_{in}$.}
    \label{fig:sens_interarea}
\end{figure}

Fig. \ref{fig:sens_interarea} shows how the dominant inter-area eigenvalue is affected by both the inertial and primary frequency response. 

As previously outlined, higher $K_p$ leads to progressively more damped modes. This is not true for the inertial response, since it can affect in a beneficial, detrimental or negligible way the inter-area mode, depending on the value of $K_{in}$. For lower values of $K_p$, the best damping is associated with $K_{in}=50$. For larger values of $K_p$, the best damping is obtained for $K_{in}=0$ (no inertial response) or $K_{in}=50$. Intermediate values of $K_{in}$ have a detrimental effect on damping in most cases. 



\begin{figure}[ht]
    \centering
    \includegraphics[width=0.5\linewidth]{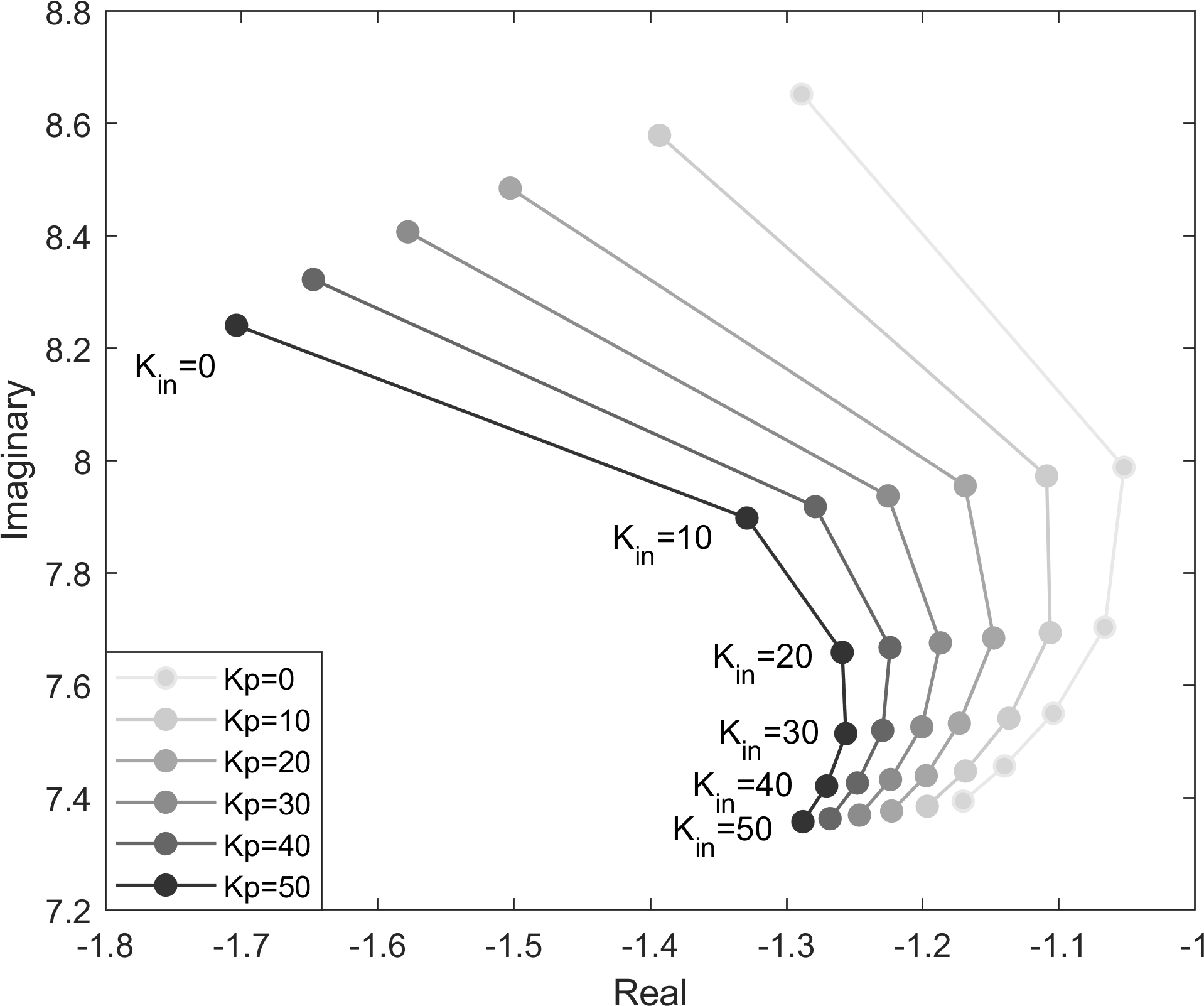}
    \caption{Sensitivity analysis of local modes to $K_p$ and $K_{in}$.}
    \label{fig:sens_local}
\end{figure}

Regarding the dominant local mode, it is very sensitive to $K_{in}$ and $K_p$ when their values are low, Fig. \ref{fig:sens_local}. However, a certain degree of saturation can be observed for large values of both. For example, the changes to damping from $K_{in}=40$ to $K_{in}=50$ are negligible, as for $K_p$. In general, the inertial response seems to have a detrimental effect on the dominant local mode. The highest damping is obtained when inertial response is lacking $K_{in}=0$ and the primary frequency support is maximum, $K_p=50$.

The converter control-based mode has an inverse behavior: its damping reduces with the increasing $K_p$. The maximum damping is reached for $K_{in}=10$ and $K_p=0$.
It is highly sensitive to inertial response provision in terms of frequency, the imaginary component changes from 15 to 37 as  $K_{in}$ varies. Electromechanical modes show a much lower sensitivity.

\begin{figure}[ht]
    \centering
    \includegraphics[width=0.5\linewidth]{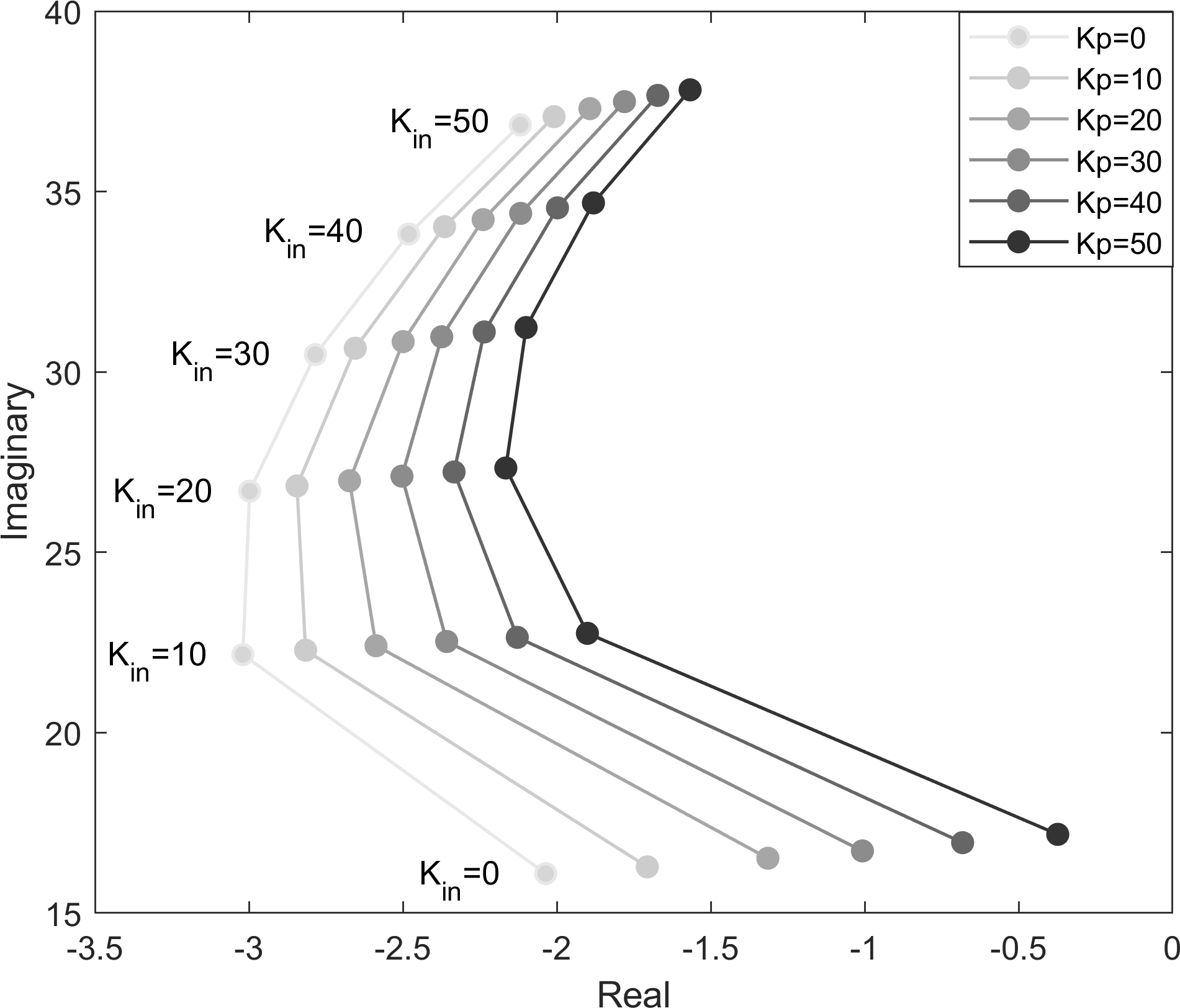}
    \caption{Sensitivity analysis of converter control-based modes to $K_p$ and $K_{in}$.}
    \label{fig:sens_converter}
\end{figure}

As a final note, the inertial response seems to affect mostly the imaginary component of the dominant modes, i.e. their frequency, whereas the primary frequency response affects the real part to a higher extent, with little influence on the imaginary one. 

These results highlight that inertial and primary frequency support provision can be beneficial to low-frequency oscillation damping. Furthermore, they stress the importance of optimal control systems design for frequency support provision required to mitigate possible associated negative impacts. 

\section{Conclusion}
In this paper, the effect of frequency support provision by DFIGs on low-frequency oscillations damping is investigated, with a discussion on both electromechanical and converter control-based modes.

The analysis was divided into three parts.
First, mathematical insights regarding the impact of primary and inertial frequency provision by a wind farm on oscillating modes were obtained for the SMIB system. 

Then, the impact of primary frequency response alone on  the dominant inter-area, local and converter control-based modes was assessed.
A small-signal stability analysis was carried out on the well-known two-area Kundur system, considering a total of nine scenarios. The resulting eigenvalue analysis was also substantiated by the results of time domain simulations. Indeed, the modes observed through the small-signal stability analysis were excited by a three-phase fault on the tie line.
The results showed that the primary frequency support provision by the wind farm has a beneficial effect on the damping of both inter-area and local modes. The largest improvements are associated with the dominant inter-area mode, and the most damped scenario corresponds to the addition of the DFIGs, providing primary frequency response, and the displacement of a synchronous generator. 
On the other hand, frequency support provision negatively affects the dominant converter control-based mode. From the results of a time domain simulation, this phenomenon does not seem to affect the stability of the whole system. However, it deserves more attention and will be explored in future work.   

Finally, a sensitivity analysis on the proportional gains $K_p$ and $K_{in}$ was conducted to assess the damping capabilities of DFIGs providing both inertial and primary frequency support. Results show that, unlike the primary frequency support, inertial response may either reduce damping or improve it, depending on the value of $K_{in}$ and on the considered mode.


These results demonstrate the importance of optimal control system design for frequency support provision, which should mitigate possible associated negative impacts and could provide an additional source of low-frequency oscillation damping.

Future works on the topic will be aimed at designing advanced and optimal controllers to provide both damping of low-frequency oscillations and frequency support, to extend these results to a real large scale system or to evaluate the impact of wind farm location.

\bibliographystyle{IEEEtran}

\bibliography{biblio_IREC}

\begin{thebibliography}{10}
\providecommand{\url}[1]{#1}
\csname url@samestyle\endcsname
\providecommand{\newblock}{\relax}
\providecommand{\bibinfo}[2]{#2}
\providecommand{\BIBentrySTDinterwordspacing}{\spaceskip=0pt\relax}
\providecommand{\BIBentryALTinterwordstretchfactor}{4}
\providecommand{\BIBentryALTinterwordspacing}{\spaceskip=\fontdimen2\font plus
\BIBentryALTinterwordstretchfactor\fontdimen3\font minus
  \fontdimen4\font\relax}
\providecommand{\BIBforeignlanguage}[2]{{%
\expandafter\ifx\csname l@#1\endcsname\relax
\typeout{** WARNING: IEEEtran.bst: No hyphenation pattern has been}%
\typeout{** loaded for the language `#1'. Using the pattern for}%
\typeout{** the default language instead.}%
\else
\language=\csname l@#1\endcsname
\fi
#2}}
\providecommand{\BIBdecl}{\relax}
\BIBdecl

\bibitem{ahmed2020grid}
S.~D. Ahmed, F.~S. Al-Ismail, M.~Shafiullah, F.~A. Al-Sulaiman, and I.~M.
  El-Amin, ``Grid integration challenges of wind energy: A review,'' \emph{IEEE
  Access}, vol.~8, pp. 10\,857--10\,878, 2020.

\bibitem{milano2018foundations}
F.~Milano, F.~D{\"o}rfler, G.~Hug, D.~J. Hill, and G.~Verbi{\v{c}},
  ``Foundations and challenges of low-inertia systems,'' in \emph{2018 Power
  Systems Computation Conference (PSCC)}.\hskip 1em plus 0.5em minus
  0.4em\relax IEEE, 2018, pp. 1--25.

\bibitem{attya2018review}
A.~Attya, J.~L. Dominguez-Garcia, and O.~Anaya-Lara, ``A review on frequency
  support provision by wind power plants: Current and future challenges,''
  \emph{Renewable and Sustainable Energy Reviews}, vol.~81, pp. 2071--2087,
  2018.

\bibitem{diaz2014participation}
F.~D{\'\i}az-Gonz{\'a}lez, M.~Hau, A.~Sumper, and O.~Gomis-Bellmunt,
  ``Participation of wind power plants in system frequency control: Review of
  grid code requirements and control methods,'' \emph{Renewable and Sustainable
  Energy Reviews}, vol.~34, pp. 551--564, 2014.

\bibitem{morren2006wind}
J.~Morren, S.~W. De~Haan, W.~L. Kling, and J.~Ferreira, ``Wind turbines
  emulating inertia and supporting primary frequency control,'' \emph{IEEE
  Transactions on Power Systems}, vol.~21, no.~1, pp. 433--434, 2006.

\bibitem{chang2009strategies}
L.-R. Chang-Chien and Y.-C. Yin, ``Strategies for operating wind power in a
  similar manner of conventional power plant,'' \emph{IEEE Transactions on
  Energy Conversion}, vol.~24, no.~4, pp. 926--934, 2009.

\bibitem{ruttledge2015emulated}
L.~Ruttledge and D.~Flynn, ``Emulated inertial response from wind turbines:
  gain scheduling and resource coordination,'' \emph{IEEE Transactions on Power
  Systems}, vol.~31, no.~5, pp. 3747--3755, 2015.

\bibitem{mahish2019distributed}
P.~Mahish and A.~K. Pradhan, ``Distributed synchronized control in grid
  integrated wind farms to improve primary frequency regulation,'' \emph{IEEE
  Transactions on Power Systems}, vol.~35, no.~1, pp. 362--373, 2019.

\bibitem{ruttledge2012frequency}
L.~Ruttledge, N.~W. Miller, J.~O'Sullivan, and D.~Flynn, ``Frequency response
  of power systems with variable speed wind turbines,'' \emph{IEEE Transactions
  on Sustainable Energy}, vol.~3, no.~4, pp. 683--691, 2012.

\bibitem{attya2017insights}
A.~B.~T. Attya and J.~L. Dominguez-Garc{\'\i}a, ``Insights on the provision of
  frequency support by wind power and the impact on energy systems,''
  \emph{IEEE Transactions on Sustainable Energy}, vol.~9, no.~2, pp. 719--728,
  2017.

\bibitem{tielens2016relevance}
P.~Tielens and D.~Van~Hertem, ``The relevance of inertia in power systems,''
  \emph{Renewable and Sustainable Energy Reviews}, vol.~55, pp. 999--1009,
  2016.

\bibitem{du2017strong}
W.~Du, X.~Chen, and H.~Wang, ``Strong dynamic interactions of grid-connected
  {DFIGs} with power systems caused by modal coupling,'' \emph{IEEE
  Transactions on Power Systems}, vol.~32, no.~6, pp. 4386--4397, 2017.

\bibitem{du2017small}
W.~Du, J.~Bi, and H.~Wang, ``Small-signal angular stability of power system as
  affected by grid-connected variable speed wind generators-a survey of recent
  representative works,'' \emph{CSEE Journal of Power and Energy Systems},
  vol.~3, no.~3, pp. 223--231, 2017.

\bibitem{gautam2009impact}
D.~Gautam, V.~Vittal, and T.~Harbour, ``Impact of increased penetration of
  {DFIG}-based wind turbine generators on transient and small signal stability
  of power systems,'' \emph{IEEE Transactions on Power Systems}, vol.~24,
  no.~3, pp. 1426--1434, 2009.

\bibitem{vittal2011rotor}
E.~Vittal, M.~O'Malley, and A.~Keane, ``Rotor angle stability with high
  penetrations of wind generation,'' \emph{IEEE Transactions on Power Systems},
  vol.~27, no.~1, pp. 353--362, 2011.

\bibitem{wilches2016impact}
F.~Wilches-Bernal, J.~H. Chow, and J.~J. Sanchez-Gasca, ``Impact of wind
  generation power electronic interface on power system inter-area
  oscillations,'' in \emph{2016 IEEE Power and Energy Society General Meeting
  (PESGM)}.\hskip 1em plus 0.5em minus 0.4em\relax IEEE, 2016, pp. 1--5.

\bibitem{du2015method}
W.~Du, J.~Bi, J.~Cao, and H.~Wang, ``A method to examine the impact of grid
  connection of the {DFIGs} on power system electromechanical oscillation
  modes,'' \emph{IEEE Transactions on Power Systems}, vol.~31, no.~5, pp.
  3775--3784, 2015.

\bibitem{allen2016measurement}
A.~J. Allen, M.~Singh, E.~Muljadi, and S.~Santoso, ``Measurement-based
  investigation of inter-and intra-area effects of wind power plant
  integration,'' \emph{International Journal of Electrical Power \& Energy
  Systems}, vol.~83, pp. 450--457, 2016.

\bibitem{dominguez2012power}
J.~L. Dom{\'\i}nguez-Garc{\'\i}a, O.~Gomis-Bellmunt, F.~D. Bianchi, and
  A.~Sumper, ``Power oscillation damping supported by wind power: A review,''
  \emph{Renewable and Sustainable Energy Reviews}, vol.~16, no.~7, pp.
  4994--5006, 2012.

\bibitem{surinkaew2014coordinated}
T.~Surinkaew and I.~Ngamroo, ``Coordinated robust control of {DFIG} wind
  turbine and {PSS} for stabilization of power oscillations considering system
  uncertainties,'' \emph{IEEE Transactions on Sustainable Energy}, vol.~5,
  no.~3, pp. 823--833, 2014.

\bibitem{dominguez2014input}
J.~L. Dom{\'\i}nguez-Garc{\'\i}a, C.~E. Ugalde-Loo, F.~Bianchi, and
  O.~Gomis-Bellmunt, ``Input--output signal selection for damping of power
  system oscillations using wind power plants,'' \emph{International Journal of
  Electrical Power \& Energy Systems}, vol.~58, pp. 75--84, 2014.

\bibitem{huang2021optimization}
J.~Huang, Z.~Yang, J.~Yu, J.~Liu, Y.~Xu, and X.~Wang, ``Optimization for {DFIG}
  fast frequency response with small-signal stability constraint,'' \emph{IEEE
  Transactions on Energy Conversion}, vol.~36, no.~3, pp. 2452--2462, 2021.

\bibitem{su2012influence}
C.~Su and Z.~Chen, ``Influence of wind plant ancillary frequency control on
  system small-signal stability,'' in \emph{2012 IEEE Power and Energy Society
  General Meeting}.\hskip 1em plus 0.5em minus 0.4em\relax IEEE, 2012, pp.
  1--8.

\bibitem{hatziargyriou2020definition}
N.~Hatziargyriou, J.~Milanovic, C.~Rahmann, V.~Ajjarapu, C.~Canizares,
  I.~Erlich, D.~Hill, I.~Hiskens, I.~Kamwa, B.~Pal \emph{et~al.}, ``Definition
  and classification of power system stability revisited \& extended,''
  \emph{IEEE Transactions on Power Systems}, vol.~36, no.~4, pp. 3271--3281,
  2020.

\bibitem{oscillations}
G.~Rogers, \emph{Power system oscillations}.\hskip 1em plus 0.5em minus
  0.4em\relax Springer Science \& Business Media, 2012.

\bibitem{quintero2014impact}
J.~Quintero, V.~Vittal, G.~T. Heydt, and H.~Zhang, ``The impact of increased
  penetration of converter control-based generators on power system modes of
  oscillation,'' \emph{IEEE Transactions on Power Systems}, vol.~29, no.~5, pp.
  2248--2256, 2014.

\bibitem{slootweg2003general}
J.~Slootweg, S.~De~Haan, H.~Polinder, and W.~Kling, ``General model for
  representing variable speed wind turbines in power system dynamics
  simulations,'' \emph{IEEE Transactions on Power Systems}, vol.~18, no.~1, pp.
  144--151, 2003.

\bibitem{van2015droop}
J.~Van~de Vyver, J.~D. De~Kooning, B.~Meersman, L.~Vandevelde, and T.~L.
  Vandoorn, ``Droop control as an alternative inertial response strategy for
  the synthetic inertia on wind turbines,'' \emph{IEEE Transactions on Power
  Systems}, vol.~31, no.~2, pp. 1129--1138, 2015.

\bibitem{keung2008kinetic}
P.-K. Keung, P.~Li, H.~Banakar, and B.~T. Ooi, ``Kinetic energy of wind-turbine
  generators for system frequency support,'' \emph{IEEE Transactions on Power
  Systems}, vol.~24, no.~1, pp. 279--287, 2008.

\bibitem{attya2019novel}
A.~B.~T. Attya and J.~L. Dom{\i}nguez-Garc{\i}a, ``A novel method to valorize
  frequency support procurement by wind power plants,'' \emph{IEEE Transactions
  on Sustainable Energy}, vol.~11, no.~1, pp. 239--249, 2019.

\bibitem{choi2016hybrid}
J.~W. Choi, S.~Y. Heo, and M.~K. Kim, ``Hybrid operation strategy of wind
  energy storage system for power grid frequency regulation,'' \emph{IET
  Generation, Transmission \& Distribution}, vol.~10, no.~3, pp. 736--749,
  2016.

\bibitem{kundur2007power}
P.~Kundur, \emph{Power system stability and control}.\hskip 1em plus 0.5em
  minus 0.4em\relax CRC press New York, NY, USA, 2007.

\bibitem{slootweg2003aggregated}
J.~Slootweg and W.~Kling, ``Aggregated modelling of wind parks in power system
  dynamics simulations,'' in \emph{2003 IEEE Bologna Power Tech Conference
  Proceedings}, vol.~3.\hskip 1em plus 0.5em minus 0.4em\relax IEEE, 2003, pp.
  1--6.

\end{thebibliography}

\end{document}